\documentstyle[aps]{revtex}
\begin{document}
\draft
\title{On the conical refraction of hydromagnetic waves in plasma
with anisotropic thermal pressure: general consideration}
\author{David Tsiklauri \footnote{email: TSIKLAURI@physci.uct.ac.za
\newline
On leave from the Department of Theoretical
Astrophysics, Abastumani  Astrophysical
Observatory, $2^{a}$ Kazbegi ave., Tbilisi 380060, Republic of Georgia.}}
\address{
Physics Department, University of
Cape Town, Rondebosch 7700, Republic of South Africa}
\date{\today}
\maketitle
\begin{abstract}
A phenomenon analogous to the conical refraction well-known in
the crystalooptics and crystaloacoustics is considered for the
magnetohydrodynamical waves in a collisionless plasma with
anisotropic thermal pressure. Imposing the most general
condition for the existence of the phenomenon,
angle of the conical refraction is
calculated which appeared to be dependent on the ratio of the
Alfven velocity and sound speed measured in the perpendicular
direction in respect to the external magnetic field.
Feasible ways of experimental demonstration of the phenomenon are discussed
and a novelty brought by the general consideration is outlined.
\end{abstract}


\section{introduction}
Recently, author of Ref.[1] showed, theoretically, that in the magnetized
plasma with anisotropic thermal pressure [viz., a medium where linear
magnetohydrodynamic (MHD) waves are
described by well-known double
adiabatic theory of Chew, Goldberger and Low [2]] there
is a possibility of observation of the
phenomenon of the conical refraction. The similar kind of phenomenon
was observed long time ago (see, for example literature in Ref.[1]) for
the light and acoustical waves in the biaxial
crystals and therefore, it is very well established. For instance,
in the biaxial crystals  observation
of the optical conical refraction is possible if the
parallel, narrow beam of natural (non-polarized) light falls 
along the optical axis (binormal) of wave normals on a plate cut so
that its two parallel faces are
perpendicular to the binormal. While propagating through the crystal
the beam will form an empty cone with permanently changing linear
polarization. Thus, producing a circle on the photographic plate
placed after the crystal.

In the Ref.[1] there were obtained the necessary conditions
which ensure existence of the conical refraction in the
medium under consideration. However, further investigation revealed
that it is possible to impose much less restrictive conditions fulfillment
of which ensures occurrence of the phenomenon. In the light of possible
interest  which plasma physics community may have to corroborate
experimentally existence of the phenomenon, author find it worthwhile
to outline here these novel conditions.

In the section II we derive the basic equations and outline new, general
conditions for the existence of the phenomenon. While in the section III
we pay more emphasis to the physics of the effect;  discuss possible
ways of its experimental corroboration along with its feasible
applications for plasma diagnostics; and finally we address a
novelty brought by the general consideration employed in the present
paper.

\section{main consideration}

Let us consider the most general form of the dispersion relation for the
MHD waves in magnetized plasma with anisotropic thermal pressure:
$$
\omega^2={{k^2}\over{2\rho_0}}
\Biggl[{{B_0^2}\over{4\pi}}+p_{\perp}(1+ \delta \sin^2{\theta})+
2 \alpha p_{\parallel} \cos^2{\theta}
$$
$$
{\pm}{\sqrt{{\left({{{B_0^2}\over{4\pi}}+p_{\perp}(1
+ \delta \sin^2{\theta})-
2 (\alpha +1) p_{\parallel}\cos^2{\theta}}\right)}^2+
4p_{\perp}^2\sin^2{\theta}\cos^2{\theta}}}\Biggr]. \eqno (1)
$$
The latter equation  can be derived from the dispersion relation obtained in
Ref.[3] while considering its  nonrelativistic limit (i.e. neglecting
displacement current and ultrarelativistic temperature effects), but retaining
a freedom in choice of the polytropic indexes $\delta$ and $\alpha$.
Note, that in the derivation of Eq.(1) there were used following
equations of state:
$$
p_{\perp}={\rm const}_\perp {\rho}B^{\delta}, \eqno(2)
$$
$$
p_{\parallel}={\rm const}_\parallel {\rho}{\left({{{\rho}}\over
{B}}\right)}^{2\alpha}. \eqno(3)
$$
We gather from the Eqs.(2) and (3) that when
$p_{\parallel}=p_{\perp}=p$ and  $\alpha = \delta = 0$ the equations of
state reduce to
conventional one for isothermal, ideal gas
$P{\sim}{\rho}$. For anisotropic pressure case, when
$p_{\parallel}\not=p_{\perp}$   and $\alpha = \delta =1$
they represent usual Chew, Goldberger and Low's equations of state [2].
Of course, taking into due account the two above limiting cases
($\alpha = \delta = 0$ and $\alpha = \delta = 1$) in Eq.(1)
results in the correct expressions for the dispersion relations for the
linear
MHD waves in isotropic and anisotropic pressure plasmas accordingly,
which can be found in any textbook of plasma physics.

For the sake of convenience we use further following notations [1]:
$$
a={{B_0}\over{\sqrt{4\pi\rho_0}}}, \,\,\,
c_\perp=\sqrt{{p_\perp}\over{\rho_0}}, \,\,\,
c_\parallel=\sqrt{{p_\parallel}\over{\rho_0}}
$$
where $a$ stands for the Alfven
velocity, $c_{\perp}$ and $c_{\parallel}$ are the sound speeds measured
in perpendicular and parallel directions in respect to the uniform,
external magnetic field ${\vec B}_0$.

Now we can rewrite Eq.(1) in the following form:
$$
V^2={{1}\over{2}}\Biggl[a^2+c_{\perp}^2(1+ \delta \sin^2{\theta})+
2 \alpha c_{\parallel}^2
\cos^2{\theta}
$$
$$
{\pm}{\sqrt{{\left({a^2+c_{\perp}^2(1+ \delta \sin^2{\theta})-
2 (\alpha +1) c_{\parallel}^2\cos^2{\theta}}\right)}^2+
4c_{\perp}^4\sin^2{\theta}\cos^2{\theta}}}\Biggr] \eqno (4)
$$
where $V=\omega/k$ denotes the phase velocity of the waves.

In Ref.[1] after this step, in order get desired result (theoretical
demonstration of the existence of the conical refraction)
author required fulfillment of the
following relation between $c_{\parallel}$ and $c_{\perp}$:
$c_{\perp}={\sqrt{2}c_{\parallel}}$. Furthermore, according to Ref.[1]
final necessary condition for occurrence of the effect was obtained as:
$a=c_{\perp}={\sqrt{2}c_{\parallel}}$.
However, now let us restrict ourselves
with much more weak condition:
$$
a^2 + c_\perp ^2 - 2(\alpha +1)c_\parallel ^2=0, \;\;\;
{\rm or} \;\;\; c_\parallel^2={{a^2 + c_\perp ^2}\over{2(\alpha +1)}}. \eqno(5)
$$
Substituting in Eq.(4) the latter expression for $c_\parallel^2$
and introducing $\chi$ and $\psi$ as
$$
\chi^2=a^2+(1+ \delta)c_\perp ^2 \,\,\, {\rm and} \,\,\,
\psi^2={{ \alpha a^2 -[\delta (\alpha +1) - \alpha]c_\perp^2 }\over{
\alpha +1}}
$$
we can obtain:
$$
V^2={1 \over 2}{\left[{ \chi^2 + \psi^2 \cos^2 \theta \pm
\sqrt{\chi^4 +2(2c_\perp ^4 - \chi ^4) \cos^2 \theta +
(\chi ^4 - 4c_\perp ^4) \cos^4 \theta
}
}\right]} \eqno(6)
$$

Possessing the formula for the phase velocity of the wave it is
easy to derive expression for the group velocity $\vec u$ via
straightforward, but time consuming calculations:
$$
{\vec u}={{{\partial}{\omega}}\over{{\partial}{\vec k}}}=
{\vec N} V+k {{\partial V}\over{\partial {\vec k}}}=
$$
$$
V \Biggl[ {{
\left[{{\cos \theta}/{2V^4}}\right]
{\left({\psi^2 T(\theta) + (2c_\perp ^4 - \chi^4)
+(\chi^4-4c_\perp ^4) \cos^2 \theta}\right)}}\over{1-C/V^4}}{\vec A}+
$$
$$
{{
\left[{{1}/{2V^4}}\right]
\left({ (2V^2- \psi^2 \cos^2 \theta) T(\theta) - (2c_\perp ^4 - \chi^4) \cos^2 \theta
-(\chi^4-4c_\perp ^4) \cos^4 \theta}\right)}\over{1-C/V^4}}{\vec N} \Biggl]
\eqno(7)
$$
where we introduced following notations:
unit vectors ${\vec N}={\vec k}/|{\vec k}|$
and ${\vec A}={\vec B}_0/|{\vec B}_0|$; $T(\theta) {\equiv}
2V^2 - (\chi^2 + \psi^2 \cos^2 \theta)$ and
$$
C{\equiv} (2 \alpha +1)[a^2+c_\perp ^2(1+ \delta \sin^2 \theta)]
\left({{{a^2 + c_\perp ^2}\over{2(\alpha +1)}}}\right) \cos^2 \theta
$$
$$
- (2 \alpha+1) {\left( {{a^2 + c_\perp ^2}\over{2(\alpha +1)}}\right) }^2
\cos^4 \theta-
c_\perp^4 \sin ^2 \theta \cos ^2 \theta.
$$
Here we must note that in Eq.(7) we used biquadratic equation
$$
V^4 - (\chi^2 + \psi^2 \cos^2 \theta ) V^2 +C=0 ,
$$
solution of which dispersion relation (6) actually is.

Let us describe a direction of the group velocity ${\vec u}$ by an
angle $\varphi$ between ${\vec u}$ and ${\vec B}_0$ (i.e., between ${\vec u}$ 
and ${\vec A}$).  For the $\tan \varphi $ we may write then
$$
\tan{\varphi}={{{\vec u} \times {\vec A}}\over{{\vec u} \cdot {\vec A}}}=
$$
$$
 {{{  {{ (2V^2- \psi^2 \cos^2 \theta) T(\theta) -
(2c_\perp ^4 - \chi^4) \cos^2 \theta
-(\chi^4-4c_\perp ^4) \cos^4 \theta }}}   }\over{ {{ (2V^2 + \psi^2 \sin^2 \theta) T(\theta) + (2c_\perp ^4 - \chi^4) \sin^2 \theta
+(\chi^4-4c_\perp ^4) \cos^2 \theta \sin^2 \theta} }}} \tan \theta .
\eqno(8)
$$
Note, that $T(\theta) \to 0$ when $\theta \to 0$.

The latter equation determines the direction
of the group velocity ${\vec u}$ whereas its modulus can be calculated
by expression:
$$
{|{\vec u}|} ={V/\cos({\varphi}-{\theta})} \eqno (9)
$$
It is worth of remarking that the Eq.(9) is a consequence of the
following expression
${\vec u}\cdot{\vec N}=V$ which can be readily shown by multiplying
Eq.(7) scalarly by $\vec N$.

Now, it can be easily shown that the expressions for group velocity and $\tan \varphi$
[Eqs. (7) and (8) correspondingly] exhibit uncertain behaviour
as $0/0$ when $\theta \to 0$. Removing of the uncertainty
(for the sake of brevity we omit, here, several steps of this task)
results in:
$$
\tan\varphi(+0)={\lim_{\theta\to0}{\tan\varphi(\theta)}}=
{{\pm} \left({{\alpha +1}\over{2 \alpha +1}}\right)
{{1}\over{1 + {(a/c_\perp )}^2 }} } \equiv
{\pm} \left({{\alpha +1}\over{2 \alpha +1}}\right) \Gamma^2, \eqno(10)
$$
$$
{|{\vec u}(+0)|}=\sqrt{\left[1+ {\left({{\alpha+1}\over{2 \alpha+1}}\right)}^2 \Gamma^4
\right] \left({{2 \alpha+1}\over{\alpha+1}}\right)}
\sqrt{{{a^2+c_\perp^2}\over{2}}}.
 \eqno(11)
$$
In Eq.(10) sign "+" and "$-$" correspond to the slow and fast MHD waves.
 It is worthwhile to note that Eqs.(10) and (11) are, of course,
in agreement with previous results [1,4]. If we consider the
isotropic pressure
limit ($p_\parallel=p_\perp=p$, $c_\parallel=c_\perp=c$, and
$\alpha=\delta=0$) and require $a^2=c^2$ [4] then Eqs.(10) and (11) reduce to
the form:
$$
\tan\varphi(+0)=
{\pm} {1 \over 2} \;\;\; {\rm and} \;\;\;
{|{\vec u}(+0)|}={{\sqrt{5}}\over{2}}c={{\sqrt{5}}\over{2}}a
\eqno(12)
$$
as it was actually obtained in Ref.[4].
Whereas, applying the condition imposed in Ref.[1]
$a=c_{\perp}={\sqrt{2}c_{\parallel}}$ we arrive at
$$
\tan\varphi(+0)=
{\pm} {1 \over 3} \;\;\; {\rm and} \;\;\;
{|{\vec u}(+0)|}= \sqrt{5 \over 3}
c_\perp=  \sqrt{5 \over 3}
a. \eqno(13)
$$
We must admit, here,
that in the Eq.(9) for the modulus of the group
velocity
from the Ref.[1] factor of $\sqrt{{3}\over{2}}$
is missing. Apparently, substituting $V(\theta=0)=\sqrt{3 \over 2} \cdot a=
\sqrt{3 \over 2} \cdot c_\perp$ and $\cos \varphi = \sqrt{10} /3$
in the Eq.(9) (of the present paper) results in Eq.(13).
This should not cause a confusion, as it does not affect the main
result of the Ref.[1] --- theoretical demonstration of possibility
of the occurrence of the conical refraction in the anisotropic plasma
plus
calculation of the angle of the conical
refraction, which is, actually, an experimentally observable physical
quantity.
We see that the results obtained in this paper are consistent with
previous contributions [1,4]. However, it must emphasized that more general
results given by Eqs.(10) and (11) bring more diversion in the predicted
phenomenon (see below).
Besides, from technical point of view it should be much
easier to produce the plasma satisfying condition (5) obtained in this
paper, rather than condition $a=c_{\perp}={\sqrt{2}c_{\parallel}}$
requested in Ref.[1].

\section{discussion}
The above consideration shows that when the wave normal
${\vec N}$ tends to reach a direction parallel to ${\vec B}_0$
(i. e. when $\theta$ tends to zero), vector of group velocity ${\vec u}$
does not take the same direction, but its limiting position
(when $\theta\to0$) is the
surface of an empty cone with opening angle
$2{\varphi}(+0) \equiv 2\arctan[\varphi(+0)]$
 given by Eq.(10). When $\theta$ is zero exactly
(which is quite unlikely from the experimental point of view as
all waves produced in a laboratory may have only finite angle
of aperture) then
${\vec u}$ is strictly parallel to ${\vec B}_0$. The discontinuous
behavior exhibited by the vector ${\vec u}$ can be explained by means of
existence of
so called [1,4] angular point which in the case of the anisotropic plasma
revealed to exist when $a^2 + c_\perp ^2 - 2(\alpha +1)c_\parallel ^2=0 $.
Thus, when
$a^2 + c_\perp ^2 - 2(\alpha +1)c_\parallel ^2=0 $
the angular point on the surfaces formed by the phase velocities
(both fast and slow MHD waves) is located in the point of intersection
of the ones with the direction of the magnetic field  ${\vec B}_0$.
Which, in turn leads to the discontinuous behaviour of ${\vec u}$ and
causes the effect. In other words, as it was mentioned in Ref.[1],
the angular point exists when phase velocities of the slow and fast
MHD waves do coincide. We observe form Eq. (6) [again, which was obtained
from general dispersion relation (4) imposing condition (5)]
that when $\theta \to 0$ slow ("$-$") and  fast ("+") branches of phase
velocity equal to each other (viz., the square root vanishes), which results
in an uncertain behavior
of its derivative by $\vec k$ i.e. group velocity.

It is interesting to note that in differ to the
result of Ref.[1], now imposing less restrictive condition (5) the
angle of the conical refraction is not a {\it fixed} number but it
depends [via Eq.(10)] on the ratio $a^2/c_\perp^2$. Which, in turn,
depends on the plasma medium physical parameters such as density and the
magnitude of the exerted
external magnetic field ${\vec B}_0$.

As regards possible experimental
corroboration of the effect, we may state the following: consider
two collisionless plasma  media contained in a uniform external
magnetic field ${\vec B}_0$. The media are separated by means of
the insulating plane perpendicular to the ${\vec B}_0$. Let us
choose physical parameters
in the second
medium as to satisfy the condition (5). If we place a diaphragm between
the first and second media then a  narrow beam of MHD waves propagating
from the first medium towards the second one, just after passing
the diaphragm will deflect from the direction of ${\vec B}_0$ and form
an empty cone with angle of opening
$2{\varphi}(+0)$.

It is worthwhile to note that one of the possible
applications of the effect of the conical refraction may be its usage
in plasma diagnostics. The condition (5) explicitly
depends on the magnitude of external magnetic field
${\vec B}_0$. So, if in the above proposed experimental setup, in the
second medium the condition (5) is not satisfied a detector of MHD waves
placed just opposite to the diaphragm (see Fig.1)  will fairly
detect the waves. However,
with variation of the magnitude of the
exerted magnetic field we may achieve fulfillment
of the condition (5) which will result in a steep drop of the detector's
indication, as due to the effect of the conical refraction the MHD waves
will deflect from the $\theta =0$ axis and form an empty cone with
opening angle $2{\varphi}(+0)$, thus avoiding arrival
of the waves into the detector (see Fig. 2).
Since the condition (5) explicitly depends on the
external magnetic field, in the event of occurrence of the
phenomenon it must be possible to measure its magnitude if the thermodynamical
parameters (pressure and density) of the plasma are known or vice versa.
Therefore, we conclude that presumably this effect can be used in plasma
diagnostics.
%
%
%

 Besides, of a purely fundamental
interest which experimental demonstration of the phenomenon might have,
probably, this effect will find its applications in practice, since
through variation of external parameters (viz., varying the
ratio $a^2/c_\perp^2$)
it is possible to regulate opening angle of the empty
conical beam of the MHD waves.
Which means, that, for instance, with simple from
technical point of view variation of the magnitude of the
external magnetic field it is possible to regulate a direction
of propagation of the energy
carried by the MHD waves. Because, it is known that
the energy propagates along the direction of the wave's
group velocity vector.
Therefore, one should not exclude possible applications of this novel
effect (revealed by the general consideration
employed in this paper) in the plasma devices like MHD generators are.

Finally, we would like to emphasize on principal difference between
the effect predicted in this paper and the effect of the conical
refraction occuring in the optically and acoustically biaxial crystals.
First, the most evident difference is that the phenomenon can take
place in media with qualitatively different physical properties
(magnetized plasma with isotropic [4] or anisotropic [1] pressure
on the one hand and the biaxial crystals on the other).
Second, in the case of biaxial crystals, let us take for concretness
optical conical refraction, mathematical basis for theoretical
formulation of the effect constitutes Maxwell's equations
along with material equations and in addition {\it electrical} anisotropy
of the medium is assumed, i.e. such substances (crystals) are
considered whose electrical excitations depend on the direction
of the electric field [5]. Whereas in our case anisotropy in the medium
is brought by the external {\it magnetic}
field and Euler's equation for
the magnetized, perfectly conducting  plasma is used.
The only similarity of these two phenomena is that the equations
which relate the phase velocity of the waves with their
direction of propagation
in respect to the anisotropy axis (optical axis in one case and the external
magenteic field in the other)
--- Fresnel's equation of wave normals [5]
and our dispertion relation (4) (including its special subcases [1] and [4])
have the same mathematical structure. Thus, geometrical
shape of the surface  formed by the group velocity vectors
when the phenomenon
takes place is qualitatively the same --- surface of the empty cone.

\section{Acknowledgements}
This research was supported in part by the Research Scholarship of the
University of Cape Town and by South African Foundation for
Research Development (FRD).

\newpage

\begin{minipage}[t]{8cm}
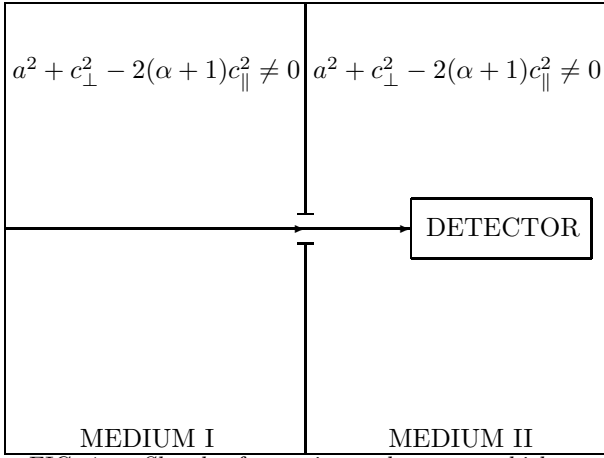
\begin{figure}
\unitlength=1mm
\begin{picture}(80,60)
\put(0,0){\line(3,0){80}}
\put(0,60){\line(3,0){80}}
\put(0,0){\line(0,0){60}}
\put(80,0){\line(0,0){60}}

\put(40,0){\line(0,0){28}}
\put(40,32){\line(0,0){28}}
\put(39,28){\line(1,0){2}}
\put(39,32){\line(1,0){2}}

\put(54,26){\line(0,0){8}}
\put(78,26){\line(0,0){8}}
\put(54,26){\line(1,0){24}}
\put(54,34){\line(1,0){24}}
\put(56,29){DETECTOR}
\put(1,50){$ a^2 + c_\perp ^2 - 2(\alpha +1)c_\parallel ^2 \not=0$}
\put(41,50){$ a^2 + c_\perp ^2 - 2(\alpha +1)c_\parallel ^2 \not=0$}
\put(10,1){MEDIUM I}
\put(51,1){MEDIUM II}

\put(0,30){\vector(3,0){40} }
\put(40,30){\vector(3,0){14} }
\end{picture}
\caption{ {\sl Sketch of experimental set up, which may be used for demonstration
of the effect of the conical refraction (see detailes in the text).
The line with arrows represents the narrow, parallel beam of MHD waves.
Note, that when in the medium II the condition
$a^2 + c_\perp ^2 - 2(\alpha +1)c_\parallel ^2=0$ is not satisfied
the beam goes into detector without experiencing any changes while
propagating through the second medium.}}
\end{figure}
\end{minipage}
\hskip 0.5cm
\begin{minipage}[t]{8cm}
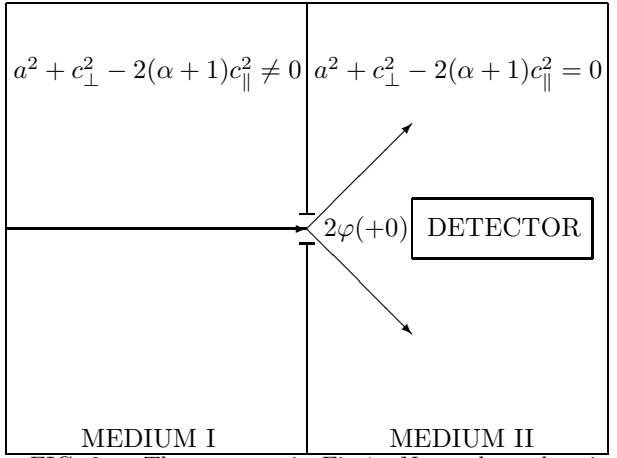
\begin{figure}
\unitlength=1mm
\begin{picture}(80,60)
\put(0,0){\line(3,0){80}}
\put(0,60){\line(3,0){80}}
\put(0,0){\line(0,0){60}}
\put(80,0){\line(0,0){60}}

\put(40,0){\line(0,0){28}}
\put(40,32){\line(0,0){28}}
\put(39,28){\line(1,0){2}}
\put(39,32){\line(1,0){2}}

\put(54,26){\line(0,0){8}}
\put(78,26){\line(0,0){8}}
\put(54,26){\line(1,0){24}}
\put(54,34){\line(1,0){24}}
\put(56,29){DETECTOR}
\put(1,50){$ a^2 + c_\perp ^2 - 2(\alpha +1)c_\parallel ^2 \not=0$}
\put(41,50){$ a^2 + c_\perp ^2 - 2(\alpha +1)c_\parallel ^2 =0$}
\put(10,1){MEDIUM I}
\put(51,1){MEDIUM II}

\put(0,30){\vector(3,0){40} }
\put(40,30){\vector(1,1){14}}
\put(40,30){\vector(1,-1){14}}
\put(42.25,29){$2\varphi(+0)$}
\end{picture}
\caption{ {\sl The same as in Fig.1.
Note, that when in the medium II the condition
$a^2 + c_\perp ^2 - 2(\alpha +1)c_\parallel ^2=0$ is fulfilled,
the beam due to the effect of the conical refraction
deflects from its initial direction ($\theta=0$) and forms an
empty cone with opening angle $2{\varphi}(+0) \equiv 2\arctan[\varphi(+0)]$,
thus avoiding arrival of the MHD waves into detector's receiver.}}
\end{figure}
\end{minipage}
\end{document}